\documentclass[fleqn,usenatbib]{mnras}

\usepackage{newtxtext,newtxmath}

\usepackage[T1]{fontenc}

\DeclareRobustCommand{\VAN}[3]{#2}
\let\VANthebibliography\thebibliography
\def\thebibliography{\DeclareRobustCommand{\VAN}[3]{##3}\VANthebibliography}

\usepackage{graphicx}	% Including figure files
\usepackage{amsmath}	% Advanced maths commands
\usepackage{multicol}
\title[NN Ser]{Investigation on the Orbital Period Variations of NN Ser: Implications for the Hypothetical Planets, the Applegate Mechanism and the Orbital Stability}

\author[A.Özdönmez et al.]{
Aykut Özdönmez,$^{1}$\thanks{E-mail: aykut.ozdonmez@atauni.edu.tr}
Huseyin Er,$^{1}$
Ilham Nasiroglu $^{1}$
\\
$^{1}$Atatürk University, Faculty of Science, Departments of Astronomy and Space Science, Yakutiye, 25240, Erzurum, Turkey\\
\\
}

\date{Accepted 2023 October 06. Received 2023 October 03; in original form 2023 May 29}

\pubyear{2023}

\begin{document}
\label{firstpage}
\pagerange{\pageref{firstpage}--\pageref{lastpage}}
\maketitle

\begin{abstract}
We present 36 new mid-eclipse times obtained between 2017 and 2023 using the T100 telescope
in Turkey, extending the time span of the $O-C$ diagram to 25 years. Once again, these new
observations show significant deviations from previous published models that were able to
explain the observed variations of the binary period. We investigate two plausible explanations
for this variability: the LTT effect due to the presence of one or two invisible low-mass
(planetary) companion(s) in distant circumbinary orbits; other mechanisms, like e.g. the
Applegate mechanism, associated with the magnetic cycles of the M-dwarf component of the
WD+dM binary. Through MCMC analyzes, we demonstrate that the observed $O-C$ variability
can be explained by the presence of a planet with a minimum mass of $\sim9.5 M_J$. This
circumbinary planet orbits around the binary system with a period of about 19.5 years,
maintaining a stable orbit for a timeline of 10 Myr. By adding a weak LTT signal from a
secondary hypothetical planet we achieve statistically better results. However, the orbits of the
bodies in a two-planet system remain stable only for a small range of the parameter space.
The energy required to power the Applegate and other Applegate-like mechanisms is too high to
explain the period variations observed. Thus, on the one hand there is substantial evidence
supporting the existence of a planet in the NN Ser system, but on the other hand there are also
compelling indications that cast doubt on the existence of a second hypothetical planet.

\end{abstract}


\section{Introduction}
\label{sec:1}
NN Ser is a well-studied short-period eclipsing pre-cataclysmic binary system containing a very hot white dwarf ($0.535$ $M_\odot$) and an M4 spectral type main-sequence dwarf ($0.111$ $M_\odot$) \citep[][and references therein]{1989Haefner, 1991ApJ...381..551W, 1994MNRAS.269..879C,  2009ApJ...706L..96Q, 2010MNRAS.402.2591P, 2016MNRAS.460.3873B}. Its brightness decreases by $\sim5.8$ mag from  $V\sim17$ mag (\textit{GAIA} $G=16.53$ mag) during the primary eclipse  with an orbital period of $\sim3.12$ hr. In addition, the system shows an orbital period variation. \citet{1989Haefner} investigated the orbital period variation of the eclipsing binary NN Ser, and updated the orbital parameters in 2004 \citep{2004A&A...428..181H}. 
\citet{2006MNRAS.365..287B} reported that the orbital period variation of the system may arise from either angular momentum loss through magnetic braking or the existence of a third object. 
The variation of the orbital period has been mostly explained by the light travel time (LTT) effect only, which results from the presence of one or two planets
\citep{2009A&A...499L...1C, 2009ApJ...706L..96Q, 2010A&A...521L..60B, 2012MNRAS.425..749H, 2013A&A...555A.133B, 2013MNRAS.436.2515M, 2016MNRAS.460.3873B}.
However, other mechanisms like magnetic cycle, magnetic breaking, gravitational radiation have been dismissed as a plausible explanation.
According to the report by \citet{2013A&A...555A.133B}, the two-planet model was dynamically stable for a lifetime greater than 100 Myr utilising the system parameters, i.e. the orbital periods of 7.65 and 15.47 years, and the minimum masses of 6.97 $M_{J}$ and 1.73 $M_{J}$ for the inner and outer planet, respectively.
In contrast to the abovementioned study, \citet{2013MNRAS.436.2515M} discussed the orbital evolution of NN Ser, and demonstrated that the orbital configuration including two planets in the binary system evolves dynamically unstable on short timescales. 
Recent studies on the orbital period variation and orbital stability suggested that the stable system configuration including two planets consistent with the $O-C$ diagram can be hypothetically constructed even for timescales larger than 1 Myr
\citep{2014MNRAS.437..475M, 2016MNRAS.460.3873B}. 
\citet{2018A&A...615A..81N} reported that the Applegate mechanism could cause additional fluctuations in eclipsing time variations.
The recent observations since 2016 are inconsistent with all existing models, contrary to the predictions of previous models in the literature \citep[see Fig 5. in ][]{2022MNRAS.514.5725P}.

\citet{2016MNRAS.459.4518H} found convincing evidence for the existence of dust around NN Ser.
Although this finding alone does not provide conclusive proof for the presence of planet(s) around NN Ser, it significantly enhances the possibility of planet formation from common-envelope material within a ’second-generation’ scenario.
In addition, a resonant interaction between the binary and the circumbinary disc may cause the loss of angular momentum leading to a slow change in the  binary's orbital period \citep{1991ApJ...370L..35A, 2011MNRAS.417.1466K, 2014A&A...562A..19V, 2015AN....336..458S, 2017ApJ...837L..19C}.

This study makes an observational contribution to the literature by presenting 36 new mid-eclipse times obtained between 2017 and 2023.
The $O-C$ diagram covers a period of 25 years by adding these eclipse times.
The details of the observations, modelling primary eclipse light curves, and the compiled mid-eclipse times are given in Section \ref{sec:2}.
This study presents a comprehensive investigation of the variation in orbital period through analysis of the $O-C$ diagram, orbital stability and the Applegate mechanism. The analysis methods and results are presented in Section \ref{sec:3}, and further discussed in Section \ref{sec:4}.

\section{Timing Data}
\label{sec:2}
We performed optical photometric observations of NN Ser between Feb 25, 2017  and Feb 18, 2023 using a SI 1100 CCD camera (field of view of $21.5\arcmin\times21.5\arcmin$ and pixel scale of 0.31 arcsec pixel$^{-1}$) with a 4 K $\times$ 4 K chip attached to the 1 m telescope of TUG T100 (TÜBİTAK National Observatory) in Antalya, Turkey. During the observations, the readout time was decreased to 2-3 seconds by using the sub-array mode ($300\times300$ pixel size) together with the binning $2\times2$ mode of the camera. Thus, the time resolution of the light curves (LC) was increased. 
We did not use any filter during the observations in order to obtain high counts within an exposure time of 5-10 seconds.
The CCD images were reduced using standard procedures, i.e. bias subtraction, flat fielding, and cosmic ray correction. Aperture photometry was performed on the reduced CCD images with the same methodology as in \citet{2021MNRAS.507..809E}.

\begin{figure}
  \includegraphics[width=\columnwidth]{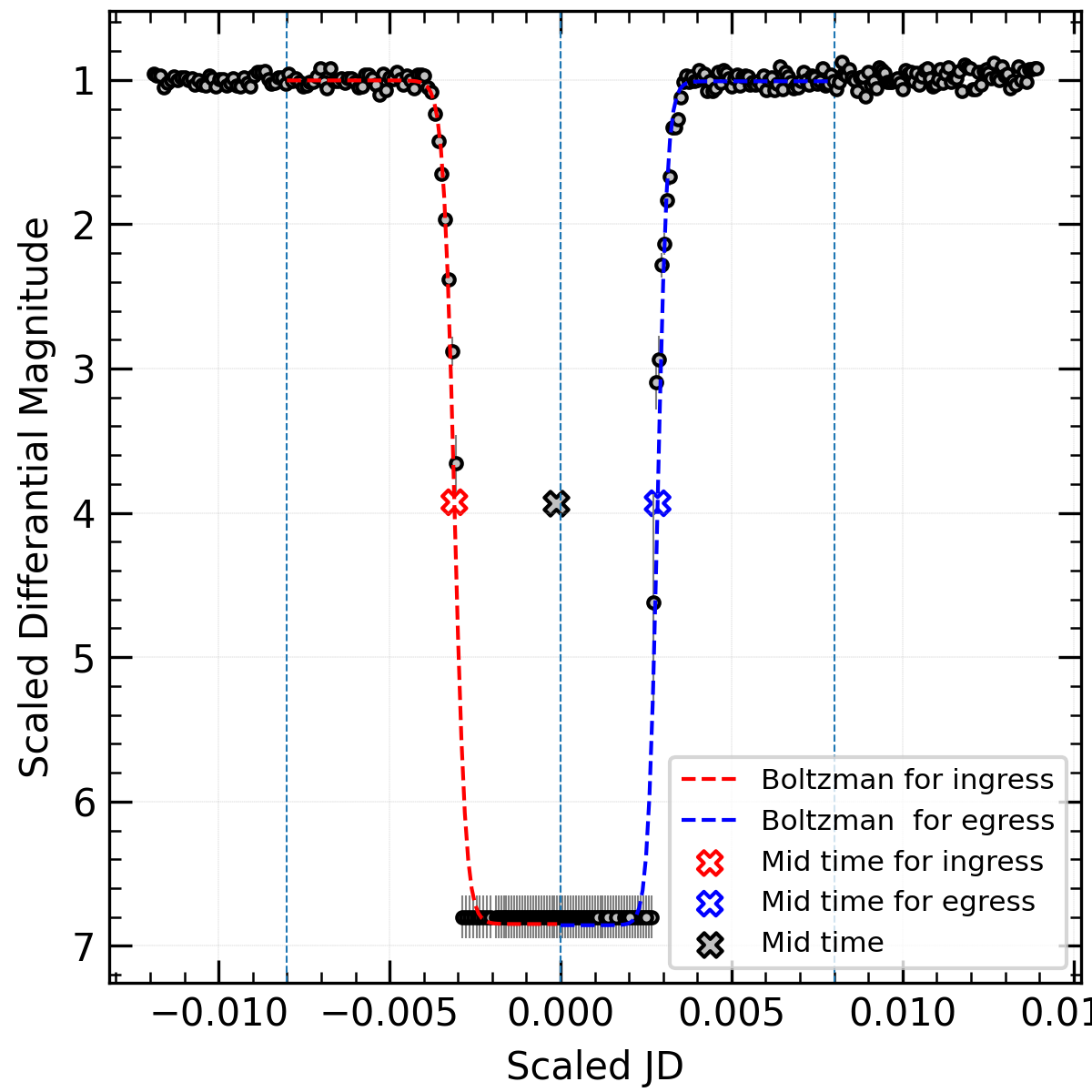}
  \caption{The light curve of NN Ser obtained from observations of TUG T100 at Feb 18, 2023 given as an example. The ingress and egress parts of the light curves were fitted individually with a Boltzmann function, as described in Sec. \ref{sec:2}. In the plot, the fits are represented by red and blue dashed lines for ingress and egress parts, respectively, and the obtained mid-times are denoted by colored "X". The vertical blue dotted lines are the limits of the ingress and egress parts of the LC.}
\label{fig:light_curve}
\end{figure}

The system becomes fainter about $5.8$ mag, reaching $\sim23$ mag during the primary eclipse \citep{2004A&A...428..181H}. The brightness at mid-eclipse is lower than the 1.0m telescope limit. 
Thus, the base magnitude of the eclipse was set to be 5.8 mag lower than the median magnitude outside the eclipse.
The errors of the base magnitudes were determined by calculating the median values of the errors observed outside of the eclipse, which were then adjusted by adding three times the standard deviation.
To determine the mid-eclipse time from the primary eclipse LC, we first divided the light curve into two parts; ingress and egress. Then, we modelled these light curve parts  with two Boltzmann functions.
As shown in Fig. \ref{fig:light_curve}, the mid-eclipse time was calculated as the mean value between the mid-ingress and mid-egress times. 
The error in the mid-eclipse time is the total errors of the uncertainties of the mid-ingress and mid-egress times acquired from the models.
Six out of 36 primary eclipse light curves do not include either  ingress or egress parts. For  these incomplete LCs, the mid-eclipse times were determined by adding half of the eclipse duration to the mid-ingress times or subtracting from the mid-egress times. The eclipse duration was calculated to be $4. 2\pm0. 35$ minutes, which represents the average time difference between the mid-times of ingress and egress for 30 complete light curves.

The mid-eclipse times presented in this study and those compiled from the literature are listed in Table \ref{tab:midtimes_nnser}. It should be noted that the mid-eclipse times obtained from the incomplete light curves are labelled as "T100-h". In Table \ref{tab:midtimes_nnser}, we present a total of 223 mid-eclipse times including 36 new primary mid-eclipse times obtained from our observations between 2017 and 2023. 
\citet{2010A&A...521L..60B} re-analyzed the mid-eclipse time calculations and made corrections to some of the values from the previous results.
We used these revised mid-eclipse times and their statistical errors from \citet{2010A&A...521L..60B}. 
On the other hand, all mid-eclipse times presented by \citet{2014MNRAS.438L..91P} are obtained from the secondary eclipse light-curves, but it was concluded that these times are not much affected by the apsidal precession, and hence they won't significantly differ from the general trend of the $O-C$ diagram. 
These mid-eclipse times weren’t used by \citet{2016MNRAS.460.3873B} in their analysis of the orbital period variation.
Besides, \citet{2016MNRAS.460.3873B} didn't list the mid-eclipse times they used, so we couldn't compile the mid-eclipse times from their observations.
For the the mid-eclipse times without any published error, we adopted a 30 s uncertainty, which is the largest data error as listed. The largest and average errors of mid-eclipse times obtained from our observations are about 10 s and 3 s, respectively. We note that the mid-eclipse times obtained from the LCs using telescopes larger or smaller than 2m diameter have approximately the same average errors.

%%%%%%%%%%%%%%%%%%%%%
\begin{table}
\caption{The Mid-eclipse Times of NN Ser, its error and references. }
\label{tab:midtimes_nnser}
\begin{tabular}{lcll}
\hline
\multicolumn{4}{c}{In literature} \\
\hline\hline 
Cycle  & BJD & $1\sigma$ error & References$^a$\\
\hline
0 & 2447344.524664 & 0.000350 & Haefner (1989)\\
2760 & 2447703.545744 & 0.000002 & Haefner et al. (2004)\\
2761 & 2447703.675833 & 0.000006 & Haefner et al. (2004)\\
2769 & 2447704.716460 & 0.000003 & Wood and Marsh (1991)\\
2776 & 2447705.627023 & 0.000003 & Wood and Marsh (1991)\\
 ... & ... & ...  & ...\\
\hline
\multicolumn{4}{c}{From our data} \\
\hline\hline 
Cycle &  BJD & $1\sigma$ error  & References\\
\hline
80451 & 2457809.602666 & 0.000018  & This work\\
80673 & 2457838.480545 & 0.000018  & This work\\
80674 & 2457838.610585 & 0.000017  & This work\\
80949 & 2457874.382660 & 0.000021  & This work\\
81456 & 2457940.333386 & 0.000019  & This work\\
... & ... & ... & ...\\
\hline
\multicolumn{4}{l}{$^{*}$ Full table is available in its entirety in machine-readable form.}\\
\end{tabular}
{\footnotesize
$^a$\citet{1989Haefner}, \citet{2004A&A...428..181H}, \citet{1991ApJ...381..551W}, \citet{2002Pigulski}, \citet{2009ApJ...706L..96Q}, \citet{2010MNRAS.407.2362P}, \citet{2010A&A...521L..60B}, \citet{2013A&A...555A.133B}, \citet{2014MNRAS.438L..91P}, \citet{2014MNRAS.437..475M}, \citet{2020JBAA..130..357F}, \citet{2022MNRAS.514.5725P} }

\end{table}
%%%%%%%%%%%%%%%%%%%%%%%%%%%%

\section{Analysis and Results} %$O-C$
\label{sec:3}
The results are presented in three sub-sections outlining the methods of analysis used; i.e. modelling the $O-C$ diagram using the Markov-Chain Monte Carlo (MCMC) technique, the orbital stability analysis by N-body integration, and searching for magnetic activity by testing the Applegate and Applegate-like mechanisms, respectively. We have investigated these topics in depth to get a better understanding of the underlying mechanisms behind the orbital period variation and the structure of the NN Ser system.

\subsection{Modelling \textit{O-C} times}
\label{sec:3_1}
The mid-eclipse time variations caused by a planet similar in size to Jovian are relatively small, a few tens of seconds.
It is expected that this variation is significantly larger than both observational and systematic timing uncertainties.
One way to minimize bias, which may be caused by a eventual scattering, is to restrict the use of timing data considering their uncertainty and the measurement method.
For this reason, two types of data sets (A and B) were used in our study similar to those used in the previous studies of NN Ser.
Data Set A (hereafter DS-A) includes all 223 mid-eclipse times available to date in the literature and our new measurements listed in Table \ref{tab:midtimes_nnser}. In Data Set B (DS-B), we excluded both the measurements having uncertainties larger than 2 seconds and those that were obtained from secondary eclipse. DS-B consists of 97 mid-eclipse times.
The average errors of mid-eclipse times for DS-A/B are 2.58 and 0.52 s, respectively.

The presence of the LTT effect can result in significant sinusoidal deviations in the orbital period of a binary system. Consequently, the linear ephemeris should be continually revised to maintain precision over time. Parameters of the linear ephemeris for DS-A are obtained as follows.

%%%%%%%%%%%%%%%%%%%%%%
\begin{equation}
\label{eq:lineer_eph_datasetA}
\begin{aligned}
\ T_{eph}(L) = & T_0 + L \times P_{bin} \\
   = & \text{BJD}\: 2447344.525240(44) + L \times 0.1300801259(8) 
\end{aligned}
\end{equation}
%%%%%%%%%%%%%%%%%%%%%%%%
Here, $T_{eph}$ is the linear ephemeris, $T_0$ is the initial ephemeris at the zero cycle (i.e. $L=0$), and  $P_{bin}$ is the orbital period of the binary system. The obtained  parameters of  linear ephemeris from DS-B have very similar values as the parameters in Eq. \ref{eq:lineer_eph_datasetA}. It also gives the same cycle (i.e. no  shift)  relative to the commonly used linear ephemeris  given by \citet{2010A&A...521L..60B}.

\begin{figure*}
\begin{multicols}{2}
\includegraphics[width=0.48\textwidth]{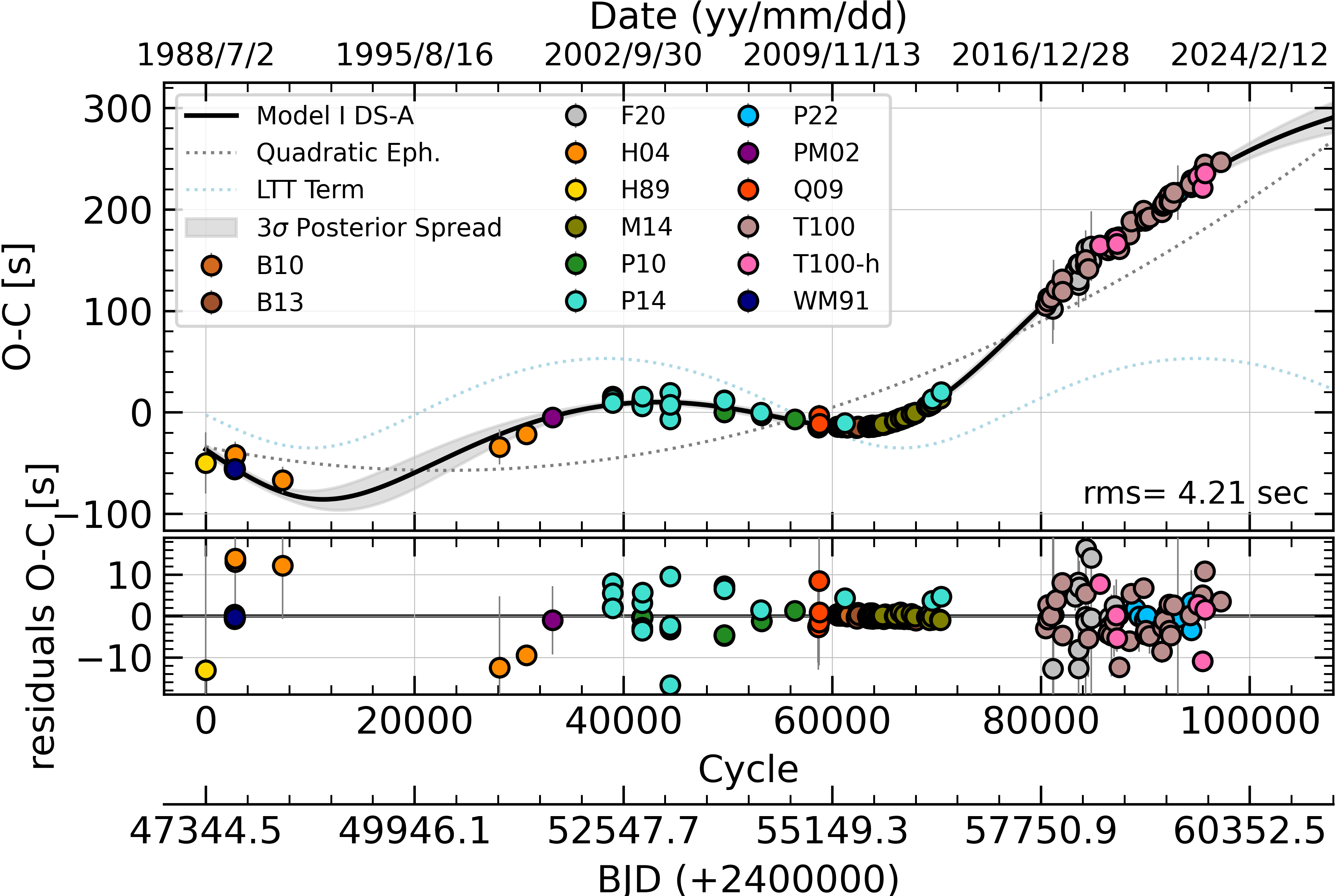}
\par
\includegraphics[width=0.48\textwidth]{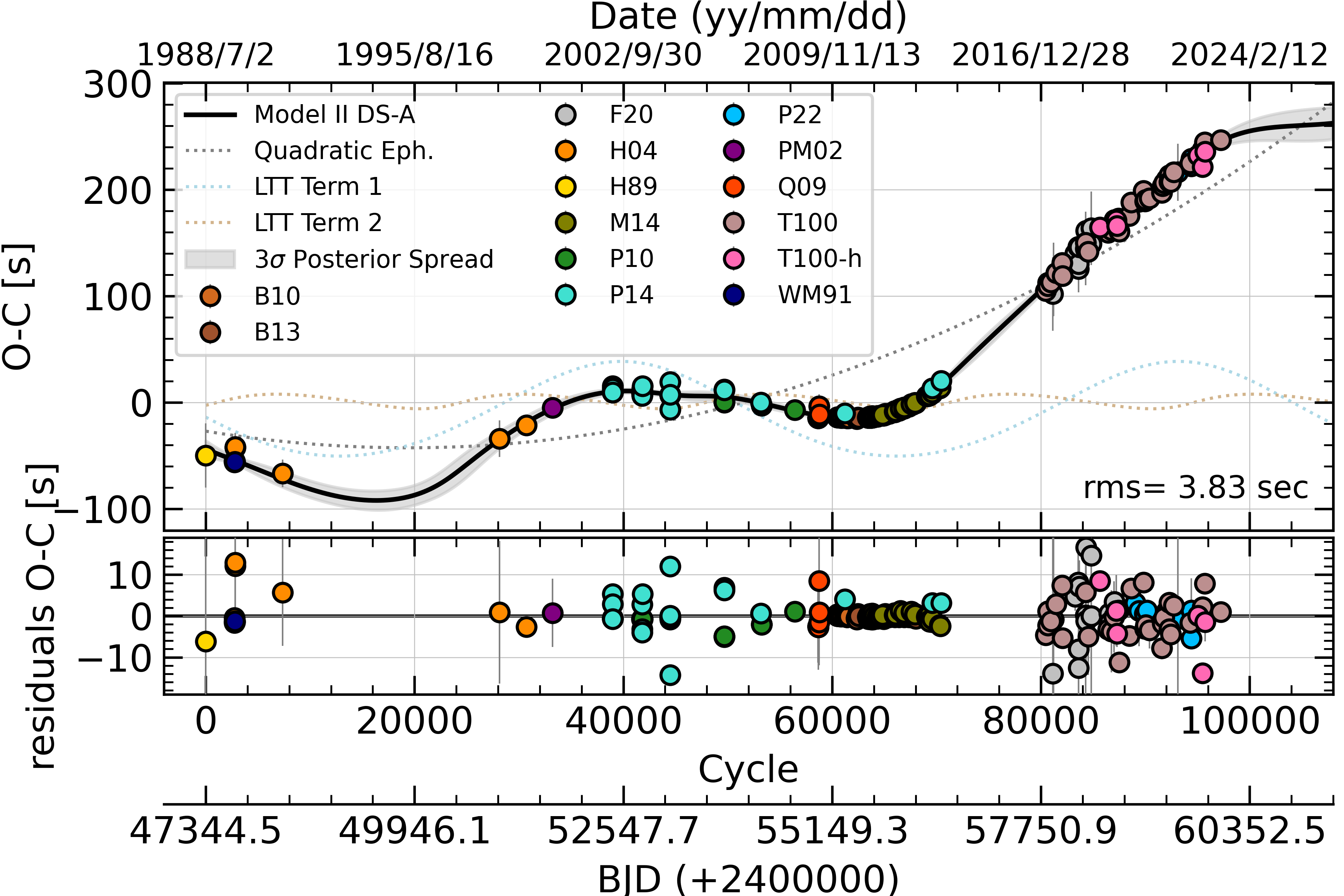}

\par
\end{multicols}
\begin{multicols}{2}
\includegraphics[width=0.48\textwidth]{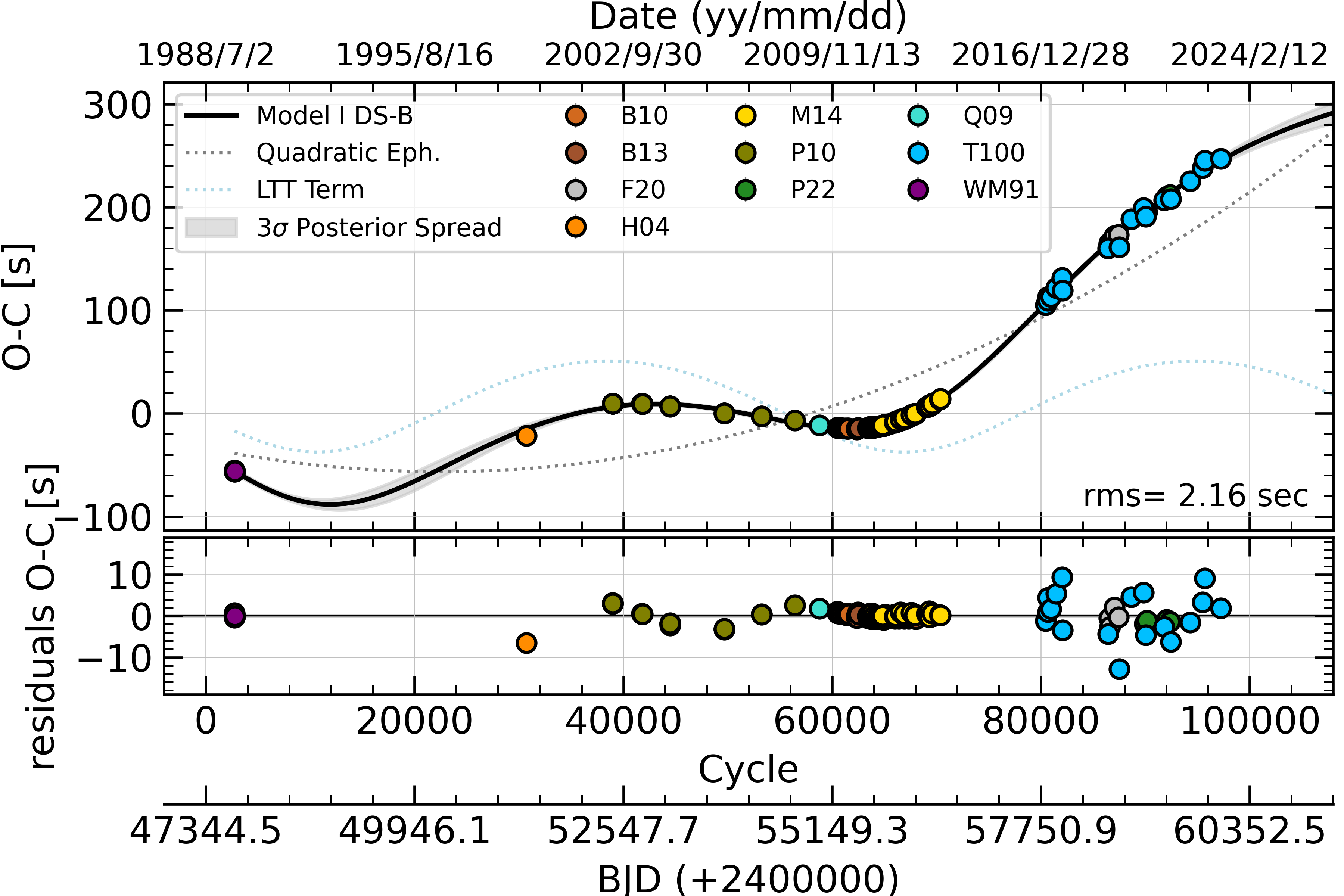}
\par
\includegraphics[width=0.48\textwidth]{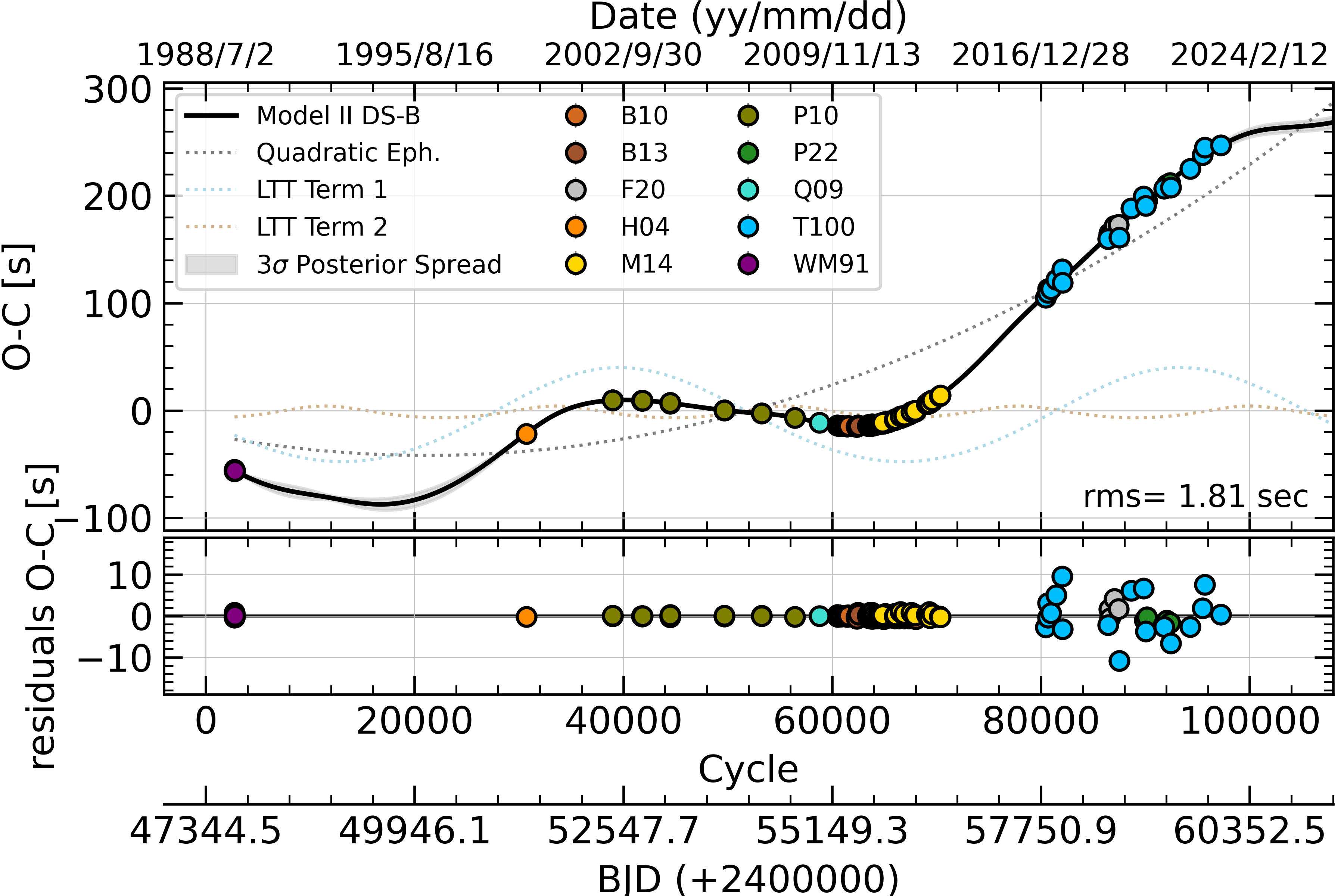}
\par
\end{multicols}
\caption{$O-C$ diagrams and their residuals for Model I (left panels) and Model II (right panels), using DS-A (upper panels) and DS-B (lower panels). Linear ephemeris times are calculated using equation \ref{eq:lineer_eph_datasetA}. 
The solid black lines indicate the best-fit model, which includes the LTT term of the hypothetical circumbinary Jovian planet(s) that is represented by colored dotted lines as well as a quadratic trend  denoted by different colored dotted lines. The shaded gray area represents the $\pm3\sigma$ posterior spread which is calculated from 1000 randomly selected parameter samples from the MCMC posterior. 
All data is represented with distinct colors and labeled with corresponding abbreviations, which serve as references; B10 \citet{2010A&A...521L..60B}; B13  \citet{2013A&A...555A.133B}; F20  \citet{2020JBAA..130..357F}; H04 \citet{2004A&A...428..181H}; H89 \citet{1989Haefner}; M14 \citet{2014MNRAS.437..475M}; P10  \citet{2010MNRAS.407.2362P}; P22 \citet{2022MNRAS.514.5725P}; PM02  \citet{2002Pigulski}; Q09 \citet{2009ApJ...706L..96Q}; WM91 \citet{1991ApJ...381..551W}. In this study, the mid-times were obtained using the T100 telescope (referred to as T100), while the times obtained from the incomplete light curves are denoted as T100-h. The RMS values are also given for each $O-C$ plot.}
\label{fig:alloc}
\end{figure*}

The $O-C$ diagrams of DS-A/B are constructed by this linear ephemeris (see Figure \ref{fig:alloc}).

To test the presence of hypothetical bodies causing the orbital period variation, we used two models:
Model I consisting of a quadratic term ($\beta$) and a LTT term ($\tau_1$) of a 3rd body for a one-planet system is described through
\begin{equation}
 \label{eq:modelI}
 T_{eph}(L) = T_{0} + L \times P_{bin} + \beta L^2 + \tau_{1}(L)
\end{equation}
Model II additionally includes an another LTT term ($\tau_2$) arising from 4th body in a two-planet system; i.e. ($\text{Model I} + \tau_2$).
The parameter  $\beta$ refers to a quadratic term which is computed as $P\dot{P}/2$, where $\dot{P}$ represents the derivative of the period over time (i.e., dP/dt).
$\tau_{i}$ represents the LTT term of the ith body as 
defined in \citet{1952ApJ...116..211I}. For the $\tau_{i}$, we used a modified formulation described by \citet{2012MNRAS.425..930G} of the form.
%%%%%%%%%%%%%%
\begin{equation}
\label{eq:tau}
 \tau_{i}= K_i \left(\sin{\omega_i} (\cos{E_i (t)} - e_i) + \sqrt{1-e_i^2} \cos{\omega_i} \sin{E_i (t)}\right)
\end{equation}
%%%%%%%%%%%%%%
Here, the semi-amplitude of the LTT signal of the $i$th body is represented by $K_i$. 
The other orbital parameters of the $i$th body are eccentricity ($e_i$), longitude of pericenter ($\omega_i$), orbital period ($P_i$), and time of pericenter passage ($t_{0,i}$).  To avoid weakly constrained $e_i$ and $\omega_i$ for quasi-circular and moderately eccentric orbits, we used Poincare elements; i.e. $x\equiv e_i cos\omega_i$ and $y\equiv e_i sin\omega_i$. The eccentric anomaly of $E_i$ includes the orbital period of the $i$th body, $P_i$, and the time of pericenter passage, $t_{0,i}$. Further information  can be found in \citet{2012MNRAS.425..930G} and \citet{2017AJ....153..137N}. 

To calculate the mass ($M_i$) of the ith body, we solved the mass function.
\begin{equation}\label{eq:mass_func}
f(M_i)=\frac{(M_i\sin{i_i})^3}{(M_{i}+M_{bin})^2}=\frac{4\pi^2(a_{12}\sin{i_i})^3}{G P_i^2}
\end{equation}
Here, $a_{12} \sin{i}_i$ is the projected semi-major axis of the binary system around the barycentre of the planetary system, and $i_i$ is the inclination of the $i$th body's orbit.
For the total mass of the binary, we adopted $M_{bin}=0.646\: M_\odot$ determined by \citet{2010MNRAS.402.2591P}. 

In order to express the eclipse time variability, we used the identical fitting process, i.e. using a Genetic algorithm (GA) and MCMC methodology based on a likelihood function ($\mathcal{L}$), as described in our previous study \citep{2021MNRAS.507..809E}. Uniform prior samples of all free parameters have been randomly assigned within the specified ranges;
i.e., $\beta, K_i, P_i, t_{0,i}, \sigma_f > 0$ days, $x_i,y_i \epsilon [-0.75,+0.75]$, $P_{bin} \epsilon [0.08, 0.15]$ days and $\Delta T_0 \epsilon [-10, +10]$. 
The free parameter $\sigma_f$ in units of days is included in the $\mathcal{L}$ function to account for systematic uncertainties ($\sigma_f$), and it scales the raw uncertainties of eclipsing times ($\sigma_i$) in quadrature; such that $\sigma^2_{i,f}\rightarrow\sigma^2_i+\sigma^2_f$ results in $\chi^2_\nu \equiv \chi^2 / \text{dof} \sim 1$.
% \citep[for more detail][]{2021MNRAS.507..809E}.
We first obtained most plausible initial parameters with GA  \citep{1995ApJS..101..309C}, and then we run MCMC to sample the posterior distribution by using the affine-invariant ensemble sampler implementation of the \textit{emcee} package \citep{2010CAMCS...5...65G} provided by \citet{2013PASP..125..306F}.
For MCMC, we used 512 initial conditions (walkers)  and assessed how these walkers behaved over 50,000 and 120,000 steps in chains of each independent variable both in Model I and II, respectively.
The best-fitting parameters and their associated uncertainties are determined by evaluating the 16th, 50th, and 84th percentiles of the marginalized distributions of maximized $\mathcal{L}$. 
The most-plausible parameters for Model I - II are listed in Table \ref{tab:results_modelI} and \ref{tab:results_modelII}, respectively. These results indicate that the bodies in the models are all Jovian-type planets due to their masses (see Table \ref{tab:results_modelI}-\ref{tab:results_modelII}).
The 1D distributions of all models utilized on both DS-A/B exhibit a distinct, prominent peak. The samples also demonstrate a relatively uniform 2D distribution centered around a particular solution. In order to conserve space, the corner plots representing the 1-D and 2-D posterior probability distributions of the system parameters sampled by MCMC are shown only for Model I using DS-A and for Model II using DS-B in Fig. \ref{fig:corner_model1} and Fig. \ref{fig:corner_model2_DSB}, respectively.

\begin{table}
\caption{The most plausible Model I and system parameters obtained by using DS-A/B, along with the corresponding RMS values of $O-C$ times.}
\label{tab:results_modelI}
\begin{tabular}{lll}
\hline\hline 
Parameters (unit) & Model I & Model I\\
  & for DS-A & for DS-B  \\
\hline
$T_0$ (BJD) & 2447344.524845(31) & 2447344.524856(19) \\
$P_{bin}$ (d) & 0.1300801024(12) & 0.1300801023(8) \\
$\beta$ ($10^{-13}$d) & $5.18\pm0.10$ & $5.23\pm0.07$ \\

$K_3$ (s) & $44.09\pm0.33$ & $44.14\pm0.23$ \\
$P_3$ (yr) & $20.18_{-0.23}^{+0.25}$ & $20.06^{+0.15}_{-0.14}$ \\
$x_3$ & $0.0015\pm0.0192$ & $0.0156\pm0.013$ \\
$y_3$ & $-0.2031\pm0.0292$ & $-0.1511\pm0.019$\\
$t_{0,3}$ (BJD) & $2456020_{-98}^{+108}$  & $2456152_{-96}^{106}$ \\

$a_{3}\sin{i_3}$ (au) & $6.43\pm0.35$ & $6.41\pm0.31$ \\
$e_3$ & $0.20\pm0.03$ & $0.15\pm0.02$\\
$\omega_3$ (deg) & $-89.56\pm0.04$ & $-84.09\pm0.52$\\
$M_3\sin{i_3}$ ($M_\text{Jup}$) & $9.43\pm0.47$ & $9.47\pm0.42$ \\
$\sigma_f$ (s) & $2.80^{+0.18}_{-0.16}$ & $1.52_{-0.11}^{+0.12}$ \\ 

RMS (s) & 4.21 & 2.16 \\
\hline
\end{tabular}
\end{table}

\begin{table}
\caption{The most plausible Model II and system parameters obtained by using DS-A/B, along with the corresponding RMS values of $O-C$ times.}
\label{tab:results_modelII}
\begin{tabular}{lll}
\hline\hline 
Parameters (unit) & Model II & Model II \\
  & for DS-A & for DS-B \\
\hline
$T_0$ (BJD)  & 2447344.524927(35) & 2447344.525244(18) \\
$P_{bin}$ (d)  & 0.1300801073(12)   &  0.1300801258(4)  \\
$\beta$ ($10^{-13}$d)   & $4.79\pm0.09$ & $5.05\pm0.04$ \\

$K_3$ (s)  & $44.52_{-0.56}^{+0.61}$  & $43.83_{-0.20}^{+0.27}$ \\
$P_3$ (yr) & $19.02\pm0.19$  & $19.13_{-0.05}^{+0.06}$ \\
$x_3$ & $-0.0180_{-0.0312}^{+0.0293}$ & $0.01975\pm0.0111$  \\
$y_3$ & $0.1343_{-0.0597}^{+0.0658}$ & $0.08728\pm0.0334$ \\
$t_{0,3}$ (BJD) & $2459591_{-281}^{+238}$ & $2459253_{-184}^{+157}$  \\

$a_{3}\sin{i_3}$ (au) & $6.19\pm0.40$ & $6.21\pm0.30$  \\
$e_3$ & $0.14\pm0.06$ & $0.089\pm0.03$ \\
$\omega_3$ (deg) & $97.63\pm1.77$ & $77.25\pm1.66$  \\
$M_3\sin{i_3}$ ($M_\text{Jup}$) & $9.91\pm0.59$ & $9.71\pm0.41$ \\

$K_4$ (s) & $7.13_{-1.41}^{+1.63}$ & $5.47\pm0.40$ \\
$P_4$ (yr) & $8.33\pm0.24$ & $7.89\pm0.14$  \\
$x_4$ &  $0.2739_{-0.1438}^{+0.1355}$  & $0.01390_{-0.0415}^{+0.0443}$ \\
$y_4$   & $-0.1220_{-0.1013}^{+0.1007}$ & $0.24022_{-0.0415}^{+0.0443}$\\
$t_{0,4}$ (BJD)  & $2453469_{-226}^{+199}$  & $2454573\pm58$  \\

$a_{4}\sin{i_4}$ (au) &  $3.57\pm3.56$ &  $3.45\pm0.81$  \\
$e_4$  & $0.30\pm0.13$ & $0.24\pm0.03$ \\
$\omega_4$ (deg)  & $-24.01\pm46.37$  & $86.68\pm0.60$ \\
$M_4\sin{i_3}$ ($M_\text{Jup}$)  & $2.76\pm1.46$  & $2.19\pm0.49$ \\

$\sigma_f$ (s) & $2.48_{-0.15}^{+0.16}$ & $0.16_{-0.03}^{+0.03}$  \\

RMS (s) & 3.83 & 1.81\\
\hline
\end{tabular}
\end{table}
%%%%%%%%%%%%%%%%%%%%%%%%%%%5

%%%%%%%%%%%%%%%%%%%%%%

\subsection{Testing Stability of Orbits}
% We evaluated the stability of the constructed orbits of NN Ser using the parameters obtained from our multi-component models. The aim is to test if the orbits remain stable for a duration compared to the expected lifetimes of the common post-envelopment phase \citep[$\sim1-100$ Myr based on the age of the system; see ][]{2007A&A...473..569H, 2012MNRAS.425..749H, 2013A&A...555A.133B, 2014MNRAS.437..475M}.

To examine the orbital stability of NN Ser for our  multi-component models,  
we used the N-body orbital integration package of \textit{REBOUND}\footnote{https://rebound.readthedocs.io} \citep{2012A&A...537A.128R} including 
a Mean Exponential Growth factor of Nearby Orbits \citep[MEGNO,][]{2000A&AS..147..205C} indicator and  a Wisdom-Holman symplectic integrator  \citep[WHFast, ][]{2015MNRAS.452..376R}. 
\textit{REBOUND} utilizes N-body integration to simulate the motion of celestial objects such as planets and stars under the influence of gravity. 
% WHFast, on the other hand, is a specific implementation of the symplectic Wisdom-Holman integrator that is the most effective when applied to systems where there is one primary body and smaller perturbations to the Keplerian orbits. 
It provides two significant results in this study: (1) The MEGNO chaotic parameter surface maps: It provides a MEGNO indicator $<Y>$ by testing the chaotic behaviour of the system components for a range of both semi-major axis and eccentricity values over a period of given time. The MEGNO indicator $<Y>$ indicates whether the orbits are chaotic or not. For the given initial conditions and time, if $<Y>$ is $\leq2$, the system will remain stable.  Values of $<Y>$ greater than 2 indicate a chaotic (unstable) orbital configuration. In such a system, the resonance term between the components is high enough to deform the orbits, but it still takes some time for the orbital parameters to diffuse to orbit-crossing values. 
If at least one particle is ejected or collides, a value of 10 is assigned to $<Y>$. 
(2) Orbital stability timeline:  It integrates the orbits for a given time, and It demonstrates the variations of the orbital parameters like the semi-major axis and eccentricity as a function of time. This can be used to determine how the planets interact, when a planet will escape the system or collide, and whether the orbit will remain stable for a given period of time.

For both simulations, 
the central binary star was treated as a single mass ($M_{tot}=0.646\: M_\odot$), and all orbits were restricted to co-planar. We set the timestep to be 0.13\% of the shortest orbital period of the inner planet, i.e. 0.01 yr \citep[see][for some discussion on the timestep choices]{2015MNRAS.452..376R}. It is assumed that the limit distance for escaping from the system is 20 AU. For these conditions and using the component parameters listed in Table \ref{tab:results_modelI} - \ref{tab:results_modelII}, we performed the dynamical stability simulations to obtain both the MEGNO values for 10 Myr and the orbital stability timeline at most 10 Myr. 

In the case of the MEGNO stability map of Model I, we integrated a one-planet system, and varied the semi-major axis $a_3$ and the eccentricity $e_3$ of the planet. For Model II, we individually evaluated $a_{3,4}$ and $e_{3,4}$ of both the outer and inner planet in a binary system with two planets. 
Our simulated dynamical result shows that the one-planet system constructed for the best-fit parameters of Models I remains stable at least 10 Myr.
In addition, both MEGNO chaos parameters using the component parameters with their errors for DS-A/B remain in the stable region,  as expected for a one-planet system. The existence of a second planet in the system (i.e. Model II) makes the interaction of the components more chaotic. 
Since the uncertainties of the system parameters determined by Model II using DS-A are very high, the result of the stability analysis predicting unstable orbital configuration is doubtful. The chaotic variation of the $a_4$ and $e_4$ parameters of the inner planet with the destabilisation of the orbit  corrupts the configuration of the system within only 2000 yr. Therefore, the plots are not shown for the abovementioned models.

\begin{figure*}
\includegraphics[width=0.7\linewidth]{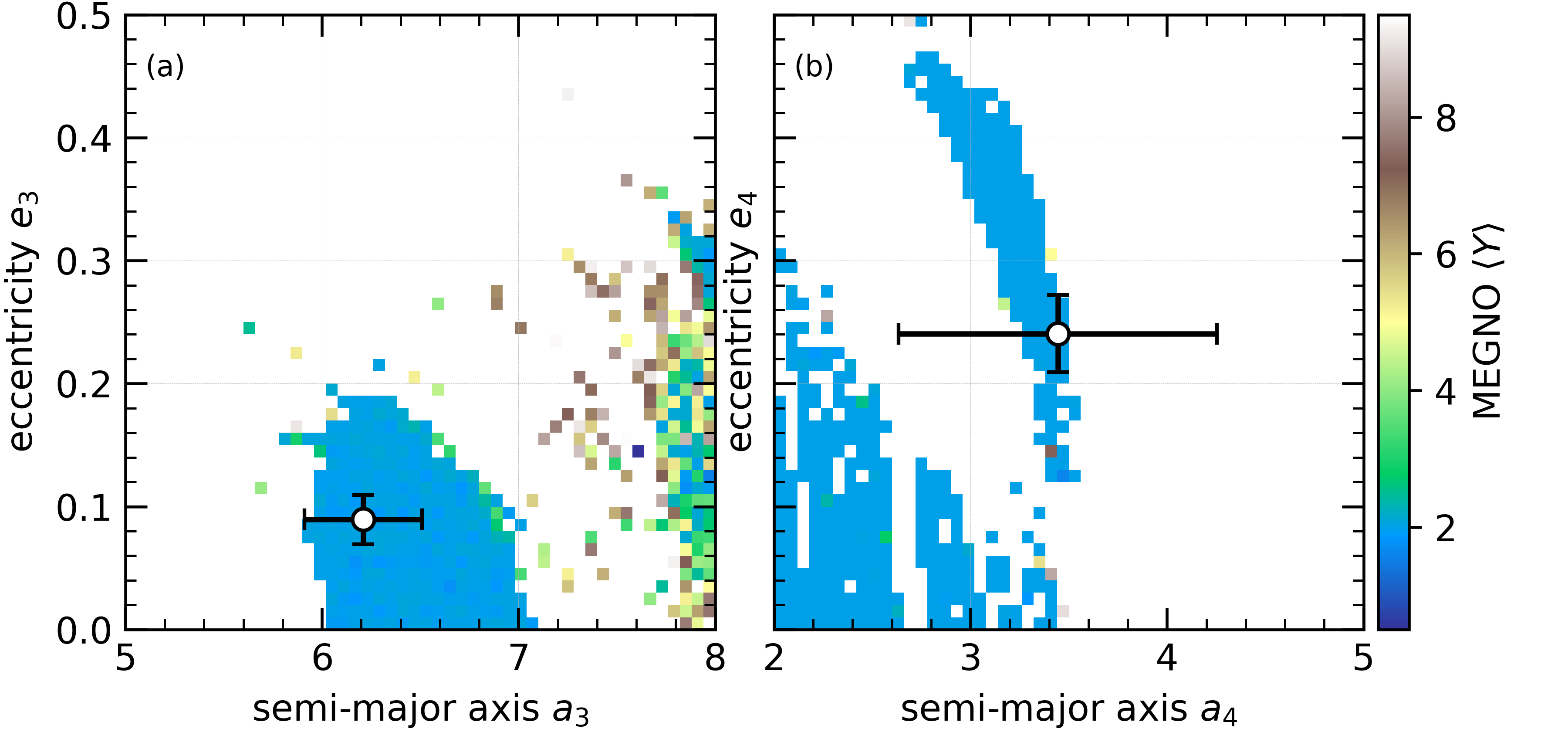}
\caption{MEGNO chaos parameter surface maps were generated for a duration of 10 Myr using the two-planet solutions from Model II DS-B, covering a range of eccentricities and semi-major axis. Specifically, the analysis was conducted for (a) the 3rd body, i.e., the outer planet, and (b) the 4th body, i.e., the inner planet. The white circles on the maps represent the best-fit model parameters, accompanied by their corresponding errors. }
\label{fig:MegnoMIIDSB}
\end{figure*}

\begin{figure}
\includegraphics[width=\columnwidth]{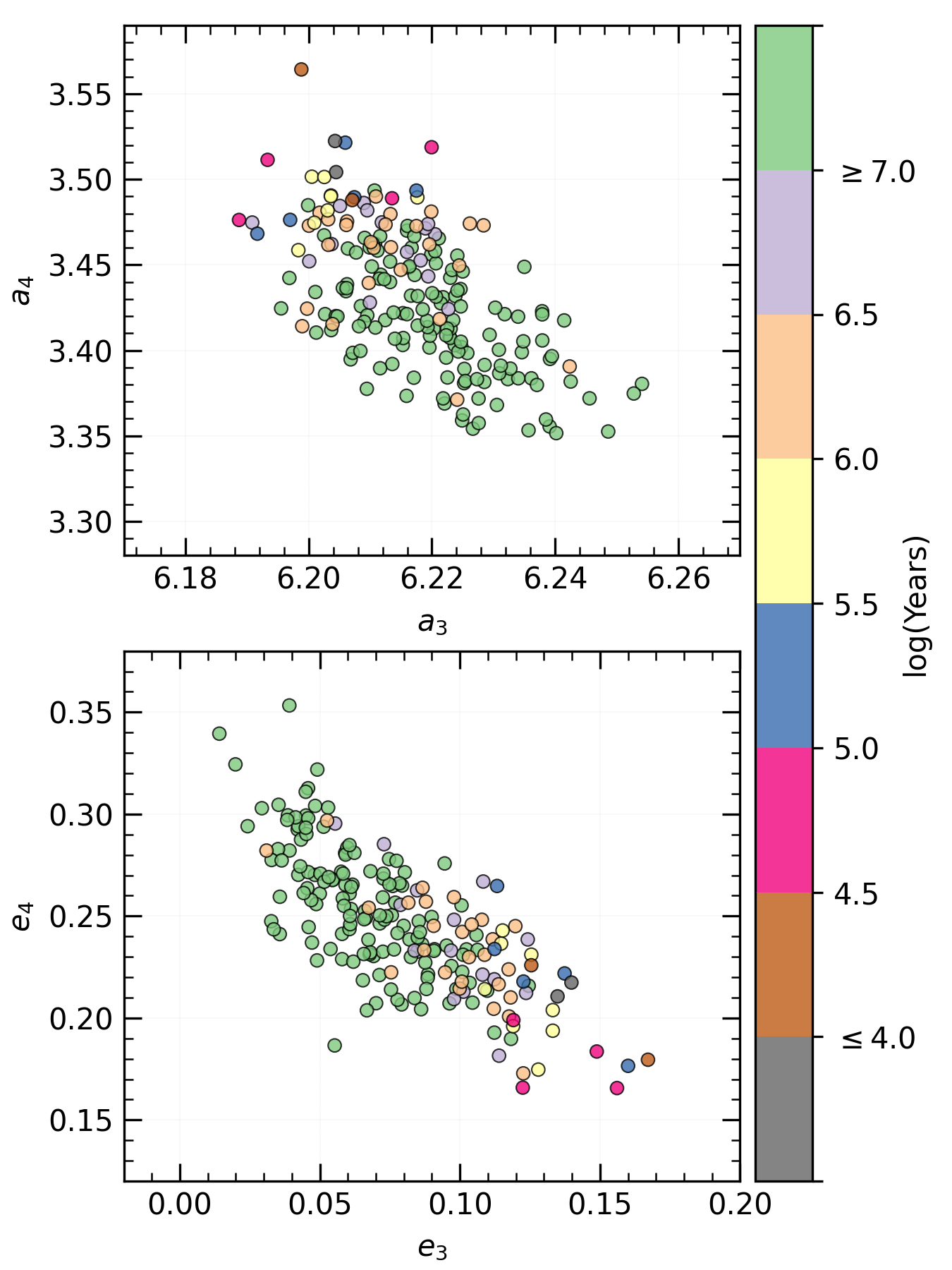}
\caption{The semi-major axis and eccentricities of the planets in the system configurations constructed
by randomly selecting 200 models from the MCMC chains of Model II DS-B. The colour bar indicates the stability time scale for each pair of values.}
\label{fig:random_whfast_a_e}
\end{figure}

For Model II DS-B, we have derived the  MEGNO stability map over a time scale of 10 Myr using the orbital solution of the best-fitting parameters (Fig. \ref{fig:MegnoMIIDSB}). In particular, the boundaries of the stable region look rather fractal.
The MEGNO chaos parameter is $\leq2$ for the best-fitting parameters of both planets, implying non-chaotic behaviour (i.e. a stable orbit). However, especially for some parameter values of the inner planet within the errors, the orbital stability is disrupted and the inner planet has escaped or collided within 10 Myr.
In order to determine the time scale on which orbits can persist undisrupted for up to 10 Myr, we obtained the orbital stability timeline of the systems  constructed by randomly selecting 200 models from the MCMC chain that also explain the $O-C$ diagram.
The distribution of both the semi-major axis and the eccentricities of the planets in these 200 systems is shown in Fig. \ref{fig:random_whfast_a_e}, considering their stability time scale. 
It reveals that 21, 36 and 5 system configurations are corrupted within 1, 1-5 and 5-10 Myr due to unstable orbits, respectively. 
However, in 138 out of 200 system configurations, the orbits remain stable over a longer time scale of more than 10 million years.
In the system configurations constructed from the MCMC chain models that are stable for more than 10 Myr, the distances between the inner and outer planets are larger, while the eccentricities of the massive outer planets, which have more influence on the resonance factor, are smaller. 

\subsection{Magnetic mechanism}
\label{sec:applegate}
The presence of substellar components in binaries may not be certain since cyclical variations of the binary orbital period may also be caused by magnetic mechanisms first proposed by \citet{1992ApJ...385..621A}. It was suggested that solar-like magnetic cycles in the secondary star can lead to changes in the angular momentum of the system, resulting in a cyclic modulation of the orbital period. 
The Applegate mechanism has been revised and improved by analyses of the magnetic mechanisms and their effects on orbital period variations by other researchers. 
\citet{2009Ap&SS.319..119T} developed an approximate equation by integrating the concepts of the Applegate mechanism and an improved model that includes changes in the quadrupole moment due to rotational and magnetic energy \citep{1998MNRAS.296..893L}. This formula establishes the relationship between the energy required to drive the magnetic mechanism and the observed eclipse time variations. 
By adopting the formulation of \citet{2006MNRAS.365..287B} in different approximations, \citet{2016A&A...587A..34V} produced the "two-zone" model, assuming different densities in the shell and core, as well as a detailed numerical model where the framework is applied to realistic stellar density profiles. This model requires far higher energy to generate the Applegate mechanism.
\citet{2020MNRAS.491.1820L} took a different approach to the magnetic mechanism by introducing a new model. This model is based on the coupling of the spin of the active star with the orbital motion of the binary system. The coupling is directly influenced by the non-axisymmetric stellar quadrupole moment, rather than by tides on much longer timescales. This cyclic exchange of angular momentum between the stellar spin and the orbit leads to the modulation of the orbital period. This proposed mechanism requires significantly less energy than previously proposed models.

We  calculated the required energy $\Delta E$ as a fraction of the available energy in the magnetically active secondary star, $E_{sec}$, to generate the corresponding magnetic mechanism from the formulations of these three studies, which have different perspectives on the magnetic mechanism. For the calculations, we used the system parameters listed in Table \ref{tab:results_modelI}-\ref{tab:results_modelII} and the astrophysical parameters of NN Ser in \citet{2016A&A...587A..34V}. It should be noted that the energy ratios have been calculated only for the substellar components that have a LTT signal with a low amplitude (see Table \ref{tab:applagate}).
In all Applegate-like models, the magnetic mechanism is sufficient to explain the orbital period variation when the ratio $\Delta E/E_{sec}<1$. 

\begin{table}
\caption{The energy ratios for the formulation of corresponding magnetic mechanisms \citep{2009Ap&SS.319..119T,2016A&A...587A..34V,2020MNRAS.491.1820L}.}
\label{tab:applagate}
\begin{tabular}{lccc}
\hline\hline 
System Parameters & Tian+ & Völschow+ & Lanza \\
for smaller LTT Signal & $\Delta E$/$E_{sec}$ & $\Delta E$/$E_{sec}$ & $\Delta E$/$E_{sec}$ \\
\hline
Model I for data set A  & 3.80 & 73.7 & 21.4 \\
Model II for data set A  & 1.41 & 25.7 & 20.3 \\
Model I for data set B  & 3.88 & 75.6 & 21.7 \\
Model II for data set B  & 0.98 & 17.6 & 17.4 \\
\hline
\end{tabular}
\end{table}

\section{Discussion and Conclusion}
\label{sec:4}

% The utilization of eclipsing timing variation has been employed to deduce the existence of sub-stellar companions orbiting in various short-period binary systems. This approach can only be verified by answering a few problems in a scientific nature, as discussed in following sub-sections. 

% \subsection{Timing Data of NN Ser}
% \label{sec:dislc}
We compiled 223 mid-eclipse times of NN Ser including 187 previously published in the literature, covering a time span of 25 years.
% This timescale is efficient to search for LTT signals in the $O-C$ diagram.}
The timescale of the periodic variation in the $O-C$ does indeed provide a clue to the mechanisms that might exist. For example, magnetic braking and gravitational radiation occur on large timescales, such as $10^8-10^9$ years, because they evolve slowly and steadily. Since the timescale of the periodic variation in the $O-C$ diagram of NN Ser is less than $\sim25$ years, it is highly unlikely that magnetic braking and gravitational radiation from the $O-C$ diagram can be detected. However, the presence of the LTT effect or the Applegate mechanism can be observed on shorter timescales of a few years to decades.

We modelled the timing variation in the most-recent $O-C$ diagram by considering quadratic ephemeris and possible LTT signals using MCMC. 
We found that the LTT signal in the one-planet model originates from a planet more massive ($M_J\sim9.4$)  than those reported in the literature,
so most of the model parameters differ from the previous studies \citep[][references therein]{2022MNRAS.514.5725P}.
The system parameters 
obtained from the solution of a one-planet system (i.e. Model I) using both DS-A/B are consistent with each other. 
However, the use of DS-B instead of DS-A for Model I reduces the systematic uncertainty from 2.80 to 1.52 seconds. 
While the root mean square (RMS) value is 4.21 seconds for the Model I obtained from DS-A,  the RMS value of 3.83 seconds is calculated for the model using DS-B. As data with large errors are removed from DS-B, both the average error of the timing data and the number of scattered data from the model are reduced, resulting in a slight decrease in the systematic uncertainty and the RMS value (see Table \ref{tab:results_modelI}-\ref{tab:results_modelII}). It can be concluded that the dataset used for the solution of the hypothetical single-planet system has no significant alteration on the results. 
This conclusion is not valid for the parameters of the two-planet solution (i.e. Model II). 
The reasons are as follows: 
1) System parameters for Model II using DS-A have very large errors. Moreover, the errors of some parameters are close to their parameter values.
2) The semi-amplitude of LTT term of the 4th body when using DS-A is  close to the RMS value of the $O-C$ and the average error of the timing data. 
3) Both the RMS value and the systematic uncertainty are not significantly reduced when the additional LTT term of the 4th body is added, especially for DS-A. The aforementioned problems are less apparent when DS-B is used to derive the parameter solution for the two-planet system. 
Therefore, the presence of the LTT signal of the 4th body should be searched only with DS-B.

Using DS-B in Model II produces the best statistical results to explain the orbital period variation of NN Ser. 
Reaching a conclusion based on statistical results alone may lead to astrophysical fault because the orbits may be highly unstable on short timescales and/or several other physical mechanisms can contribute to the timing variations.
When we analyse the orbital stability of the system, a one-planet system is stable throughout almost the entire evolutionary timeline. However, when a second planet is included into the system, the interaction between the components becomes more chaotic, amplifying the resonant terms. Based on our analyses, it has been determined that even minor alterations in the parameters of the two-planet system can disrupt its orbital stability. 
For 20.5\% of the orbital configurations formed from the MCMC chain models, orbital stability is maintained only for a duration ranging from 1 to 10 Myr. In addition, a two-planet system constructed using 69\% of the parameter solutions remains stable throughout a significantly longer time scale of 10 Myr.
These results are also consistent with the MEGNO stability map obtained for a two-planet system. Changing the parameters within their respective error ranges in the MEGNO map derived for 10 Myr causes the transition of the planetary orbits from stable to disrupted.
It is difficult to make strong statement on the orbital stability of NN Ser when only the best-fitting model is stable and most around it in the uncertainty range is unstable.

To examine the possibility of cyclical magnetic activity playing a role in generating a hypothetical signal instead of the LTT signal, we calculated the energy ratios $\Delta E/E_{sec}$ for the smaller LTT signal in our models by using three different formulations of the magnetic mechanism. Based on the formulation proposed by \citet{2009Ap&SS.319..119T}, the ratio $\Delta E/E_{sec}$ was calculated to be approximately 1 for the secondary (inner) planet in both our models DS-A/B. This value suggests that the magnetic mechanism could potentially contribute to fluctuations in the eclipsing time variations. In all the other cases, the ratio $\Delta E/E_{sec}$ was found to be significantly greater than 1 (see Table \ref{tab:applagate}). Consequently, the Applegate mechanism alone is insufficient to explain the observed variations of the binary period.
The existence of planet(s) is also supported by the detection of dust around NN Ser \citep{2016MNRAS.459.4518H}. The investigated signal of the 4th planet may also potentially be attributed to a resonant interaction between the binary system and the circumbinary disc.

Based on our models, we predict a decline phase of cyclical variation  in the $O-C$ diagram expected to occur in 2024-2025, particularly for Model II. However, the upward trend resulting from the quadratic ephemeris is expected to continue steadily for all models. To validate our predictions, further observations in the upcoming years will be essential.

\section{Acknowledgements}
This work has been supported by The Scientific and Technological Research Council of Turkey (TUBITAK), through project number 114F460 (I.N., H.E.). We would like to thank the team of the TUBITAK National Observatory (TUG) for partial support in using the T100 telescope (with project numbers TUG T100-631 and TUG T100-1333).  We would like to thank Kai Schwenzer and Ergun Ege for their help with the proofreading.

\textit{Software}: Python packages (ccdproc \citep{2017zndo...1069648C}, Astropy \citep{2013A&A...558A..33A}, Numpy \citep{2020Natur.585..357H}, Matplotlib \citet{2007CSE.....9...90H}, Photutils \citep{2020zndo...4049061B}, LMFIT \citep{lmfit}, REBOUND \citep{2012A&A...537A.128R}, emcee \citep{2013PASP..125..306F}, corner.py \citep{2016JOSS....1...24F}, Applegate calculator: http://theory-starformation-group.cl/applegate.

\section*{Data Availability}
The data underlying this article are available in the article and in its online supplementary material.



\bibliographystyle{mnras}

%%%%%%%%%%%%%%%%% APPENDICES %%%%%%%%%%%%%%%%%%%%%

\appendix
\label{sec:Appendix}
\section{Corner Plots}
\begin{figure*}
	\includegraphics[width=\textwidth]{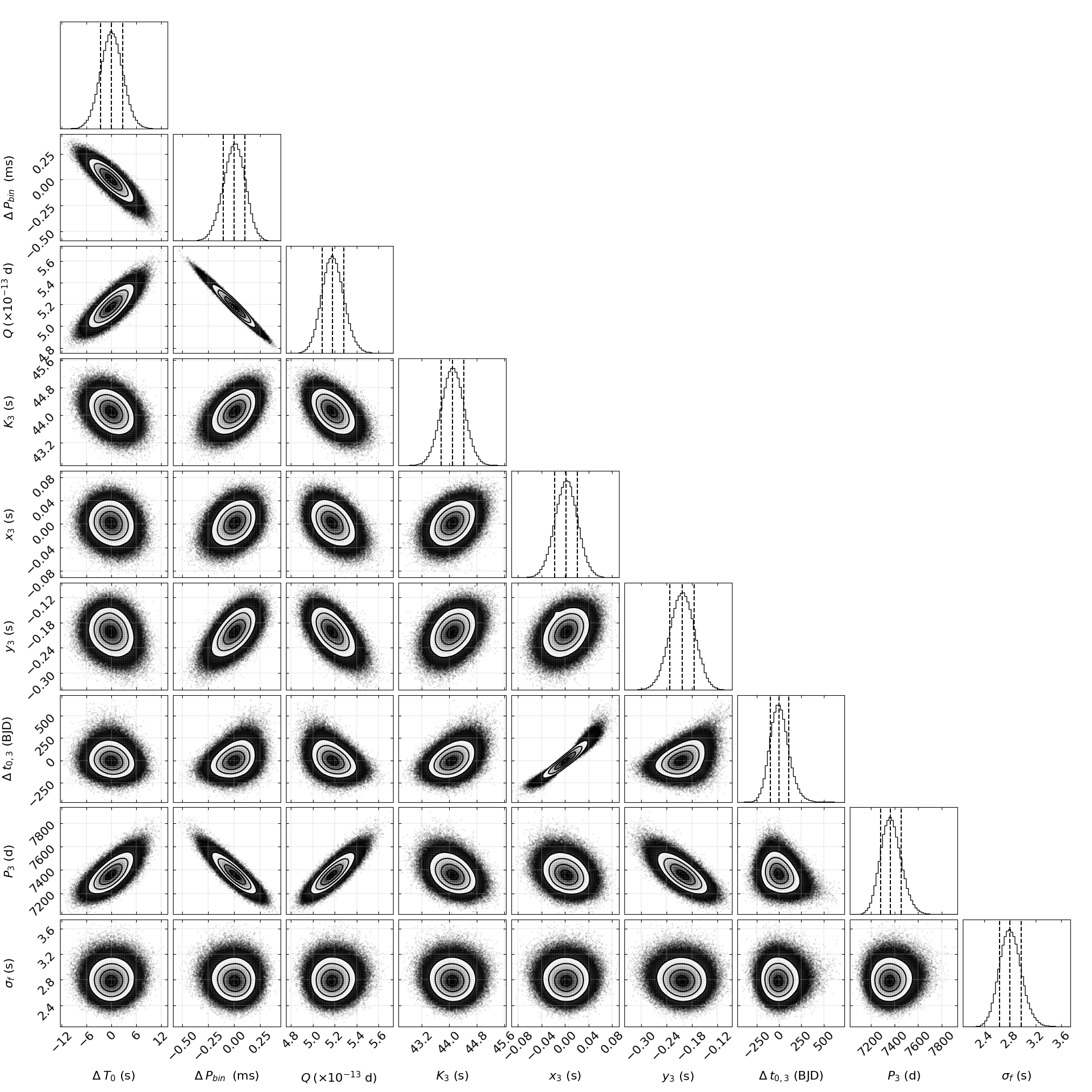}
    \caption{1-D and 2-D projections of the posterior probability distributions of the free parameters inferred from the Model I for Data set A. The samples are thinned by selecting one on every 100 samples after removing burnout sample. $\Delta$ represent differences between posterior and calculated values of the parameter. Contours are for the 16th, 50th, and 84th percentile of samples in the posterior distribution. This figure is derived using corner.py \citep{2016JOSS....1...24F}.}
    \label{fig:corner_model1}
\end{figure*}

\begin{figure*}
	\includegraphics[width=\textwidth]{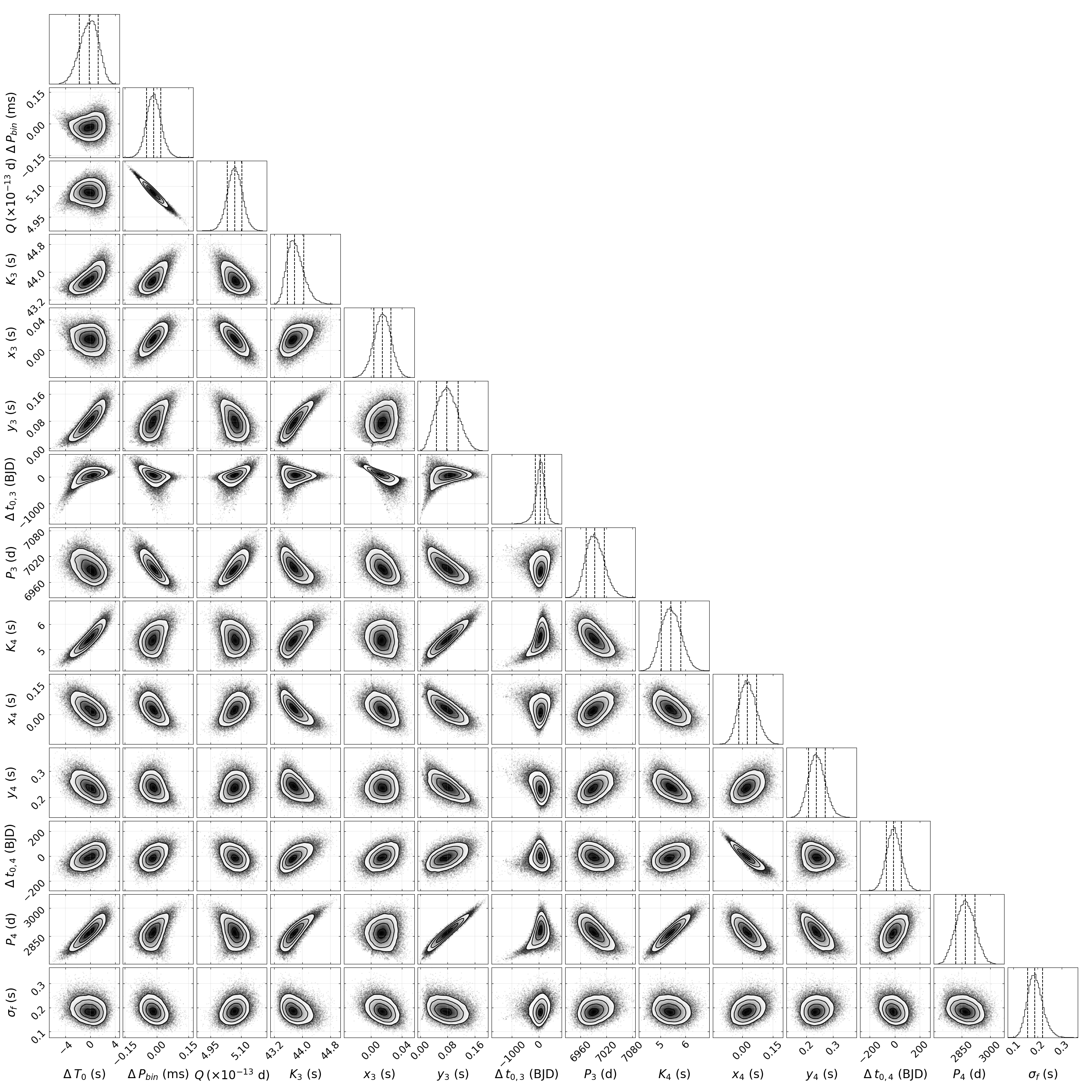}
    \caption{1-D and 2-D projections of the posterior probability distributions of the free parameters inferred from the Model II for Data set B. The other descriptions are the same as in Fig. \ref{fig:corner_model1}}
    \label{fig:corner_model2_DSB}
\end{figure*}

%%%%%%%%%%%%%%%%%%%%%%%%%%%%%%%%%%%%%%%%%%%%%%%%%%

% Don't change these lines
% \bsp	% typesetting comment
\label{lastpage}
\end{document}